\documentclass[twocolumn,showpacs,eqsecnum,preprintnumbers,amsmath,amssymb,floatfix,prb]{revtex4}
\usepackage{graphicx}
\usepackage{dcolumn}
\usepackage{bm}

\begin{document}

\title{Short-Range Order in a Flat Two-Dimensional Fermi Surface}
\author{Eberth Corr\^{e}a}
\email{eberth@iccmp.org}
\author{Hermann Freire}
\author{A. Ferraz}
 \affiliation{Laborat\'{o}rio de Supercondutividade,\\
 Centro Internacional de F\'{i}sica da Mat\'{e}ria Condensada,\\
 Universidade de Bras\'{i}lia - Bras\'{i}lia, Brazil}

\date{\today}

\begin{abstract}
We present the two-loop renormalization group (RG) calculations of
all the susceptibilities associated with the two-dimensional flat
Fermi surface with rounded corners (FS). Our approach follows our
fermionic field theory RG method presented in detail earlier on. In
one loop order our calculation reproduce the results obtained
previously by other RG schemes. All susceptibilities diverge at some
energy scale and the antiferromagnetic SDW correlations produce
indeed the dominant instability in the physical system. In contrast,
in two-loop order, for a given initial set of values of coupling
constant regime only one of the susceptibilities at a time seems to
diverge.
\end{abstract}


\maketitle

\section{Introduction}

In the last decades there have been important developments in both
experiment and theory which resulted in a better understanding of
the high temperature superconductors. One typical characteristic of
those compounds is their extreme sensitivity with doping. At
half-filling the cuprates are Mott insulators. However, at very low
hole doping, as soon as the antiferromagnetic ordering is destroyed
those materials are turned into the pseudogap
phase\cite{Damascelli,Loeser,Marshall,Ding}.

From the start Anderson\cite{Anderson} argued that the Hubbard model
(HM), which considers only short-range interactions, describes
qualitatively well the electronic properties of the high-Tc
superconductors. In the one-dimensional case (1D) this model
captures the main features displaying the existence of the Luttinger
liquid for $U>0$, spin-charge separation, as well as a Mott
insulating regime. There is no exact solution for the HM in $D>1$.
Moreover successful methods in $D=1$ such as bosonization\cite{Voit}
and the Bethe \emph{ansatz}\cite{Lieb} are either inapplicable or
simply much too hard to implement in higher dimensions. One approach
which is equally successful in both one and higher dimensions is the
renormalization group (RG) method. Different RG schemes are already
available to describe strongly interacting fermions in the presence
of a Fermi surface
(FS)\cite{Zanchi,Metzner,Honerkamp,Dzyaloshinskii,Doucot}.

In this work we continue to explore the field theoretical RG method.
We calculate the renormalized form factors of the main physical
instabilities of the two-dimensional HM up to two-loop order. We
compare our results with other RG approaches in
2D\cite{Metzner,Honerkamp}. Those results are summarized in the
phase diagram which displays all leading and subleading
instabilities in that dimension in coupling constant space. To
reinforce our case we test our method in well known grounds and show
it reproduces the analogous phase diagram in the one-dimensional
case\cite{Solyom}. In 1D the coupling functions approach fixed point
values producing susceptibilities which exhibit power-law behavior.
This reflects the absence of spontaneous symmetry breaking and
long-range order. We must keep this in mind when analyzing the
possible existing states in the phase diagrams of such low
dimensional systems. Here in specifying such phase diagrams we find
convenient to define appropriately symmetrized response functions
with respect the spin projections of the corresponding
particle-particle and particle-hole operators. In this way, we
calculate the corresponding charge density wave $(CDW)$, the spin
density wave $(SDW)$, the singlet superconductivity $(SSC)$, and the
triplet superconductivity $(TSC)$ one-particle irreducible functions
$\Gamma _{R}^{(2,1)}$'s which in turn define the respective
susceptibilities.

In a 1D metal the FS reduces to two Fermi points $(\pm k_{F})$. In
2D, for non-interacting electrons, at half-filling the FS is a
perfect square. If we go slightly away from half-filling, doping the
system with holes, all corners get rounded and the area inside the
FS is reduced from its square value. As a result in such conditions
we may neglect the van Hove singularities as well as the Umklapp
effects. In this 2D scenario a new symmetry arises with respect to
the sign of momentum component along the FS. This will reflect
itself in the RG equations for the renormalized form factors which
we redefine as the charge density wave of s and d-types $(CDW\pm)$,
the spin density wave of s and d-types $(SDW\pm)$, the singlet
superconductivity of s and d-types $(SSC\pm)$ and, finally, the
triplet superconductivity of s and d-types $(TSC\pm)$. All those
symmetries are derived from the particle-particle and particle-hole
operators which have a transfer momentum $\mathbf{q}$. Our results
for the uniform spin and charge susceptibilities were presented in a
separated publication\cite{isl}.

We showed in another work\cite{nosso} that for such flat FS as soon
as we add interactions to the 2D electron gas the flow to strong
coupling destroys the Landau Fermi liquid regime. We demonstrated
how the self-energy effects lead to the nullification of the
quasiparticle weight directly affecting the RG equations for the
renormalized coupling functions. In this work we discuss how those
results also influence directly the response functions of the
physical system. In particular we reveal that for the lightly doped
2D HM, all susceptibilities flow to fixed values. We argue that such
a behavior suggests the existence of short-range correlations
typical of a spin liquid phase. This is in agreement with our
already mentioned uniform susceptibilities results\cite{isl}.

The presentation of our results begins with the derivation of the
corresponding RG equations for the vertex functions and the
associated susceptibilities within the field theory method. We
discuss initially one-loop order results and show that they
reproduce both the parquet and other fermionic RG schemes. Following
this we present our new two-loop results. We show that the SDW and
the d-wave superconductivity are indeed the leading instabilities in
the physical system when the parallel component of the transfer
momentum $q_{\parallel}$ is equal to zero. Finally we conclude our
work discussing the physical meaning of our results.

\section{The Model}

\begin{figure}[t]
  \includegraphics[height=2.5in]{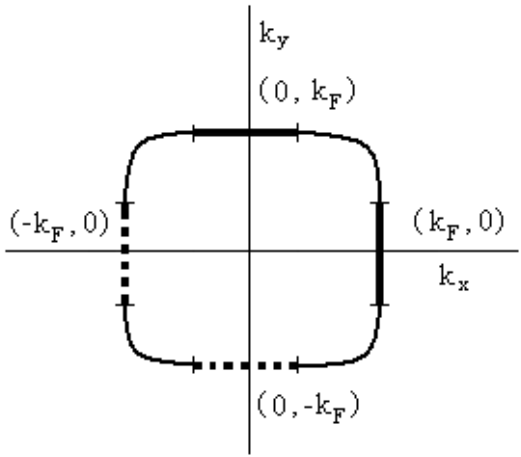}\\
  \caption{The 2D flat Fermi surface with rounded corners. We divide it into
  four regions: two of the solid line type and two of the dashed line type.
  The perpendicular regions do not mix in our scheme.}\label{sf}
\end{figure}

Firstly we consider a 2D lightly doped FS with no van Hove
singularity as shown in Fig.\ref{sf}. For convenience and to keep a
closer contact with well-known works in one-dimensional
physics\cite{Solyom}, we divide the FS into four regions. We
restrict the momenta at the FS to the flat parts only. The
contributions which originate in scattering processes associated
with FS patches which are perpendicular to each other are irrelevant
in the RG sense. For this reason we can select, for example, the
patches centered at $(0,k_{F})$ and $(0,-k_{F})$ and simply neglect
the other two. Following this we will restrict ourselves to the
one-electron states labeled by the momenta $p_{\perp}=k_{y}$ and
$p_{\parallel}=k_{x}$. Accordingly the momenta parallel to the FS is
restricted to the interval $-\Delta\leqslant
p_{\parallel}\leqslant\Delta$, with $\Delta$ being essentially the
size of the flat patches. In the same way the energy dispersion of
the single-particle states is one-dimensional and given by
$\varepsilon_{a}\left(\mathbf{p}\right)=v_{F}\left(\left|p_{\perp}\right|-k_{F}\right)$
where $v_{F}$ is the Fermi velocity and the energy $\varepsilon$ is
measured with respect the chemical potential $\mu$. Notice that this
dispersion relation depends only on the momenta perpendicular to the
Fermi surface, where the label $a=\pm$ refers to the flat sectors at
$p_{\perp}=\pm k_{F}$ respectively. In doing so we are only
considering the flat parts whose nesting vectors are
$\mathbf{Q}^{*}=(0,\pm 2k_{F})$. We consider a fixed momentum cutoff
$\lambda$ which results in the interval
$k_{F}-\lambda\leqslant\left|p_{\perp}\right|\leqslant
k_{F}+\lambda$ for the perpendicular component in the vicinity of
the FS. We neglect how the interactions renormalize the FS itself
since it would be too complex to do otherwise at this stage. As a
result, the momentum $k_{F}$ will not be renormalized in our
approach and will resume its noninteracting value. For the same
reason, we will also neglect the Fermi velocity $v_{F}$ momentum
dependence along the FS.

Since we showed in detail all the necessary steps to do the
renormalization in such a model up to two-loop order, from the
start, we rewrite the renormalized Lagrangian $L$ associated with
the 2D flat Fermi surface entirely in terms of renormalized fields
and couplings

\begin{align}
&L=\sum_{\mathbf{p},\sigma,a=\pm}Z\psi_{R(a)\sigma}^{\dagger}\left(\mathbf{p}\right)
\left[i\partial_{t}-v_{F}\left(|p_{\perp}|-k_{F}\right)\right]\psi_{R(a)\sigma}\left(\mathbf{p}\right)\nonumber
\\&-\frac{1}{V}\sum_{\mathbf{p,q,k}}\sum_{\alpha,\beta,\delta,\gamma}\left[\prod_{i=1}^{4}Z\left(p_{i\parallel}\right)\right]^{\frac{1}{2}}\left[
g_{2B}\delta_{\alpha\delta}\delta_{\beta\gamma}-
g_{1B}\delta_{\alpha\gamma}\delta_{\beta\delta}\right]\nonumber\\
&\times\psi_{R\left(+\right)\delta}^{\dagger}\left(\mathbf{p+q-k}\right)\psi_{R\left(-\right)\gamma}^{\dagger}\left(\mathbf{k}\right)
\psi_{R\left(-\right)\beta}\left(\mathbf{q}\right)\psi_{R\left(+\right)\alpha}\left(\mathbf{p}\right),
\label{lagr}
\end{align}

\begin{figure}[b]
  \includegraphics[width=2.8in]{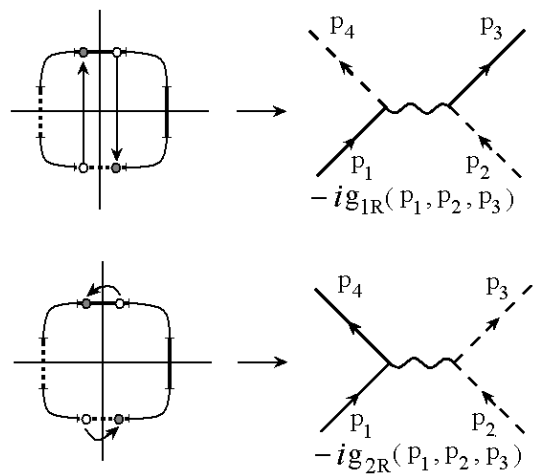}\\
  \caption{The interaction processes in the model and the corresponding Feynman rules for the vertices.
  The $g_{1R}$ and $g_{2R}$ couplings stand for the renormalized backscattering and forward
  scattering respectively.}\label{processes}
\end{figure}

\noindent following the ``g-ology'' notation. Here the
$\psi_{R\left(\pm\right)}^{\dagger}$ and $\psi_{R\left(\pm\right)}$
are, respectively, the creation and annihilation operators for
particles located at the $\pm$ patches. The couplings $g_{1B}$ and
$g_{2B}$ stand for \emph{bare} backscattering and forward scattering
couplings. From now on we will consider the thermodynamic limit
$(V\rightarrow\infty)$ with all momenta summations becoming
integrals such as $\sum_{\mathbf{p}}\rightarrow V\int
d^{2}\mathbf{p}/(2\pi)^{2}$, with $\hbar=1$. Finally, the
\emph{bare} coupling functions are related to their renormalized
associates by

\begin{equation}
g_{iB}=\left[\prod_{i=1}^{4}Z\left(p_{i\parallel}\right)\right]^{-\frac{1}{2}}\left(g_{iR}+\Delta
g_{iR}\right).\label{coup}
\end{equation}

The diagrammatic representations of the corresponding renormalized
forward and backscattering interactions are shown schematically in
Fig.2. Here, we do not consider the \emph{Umklapp} processes due to
the fact that there is no superposition between our FS and the
correspondingly \emph{Umklapp} surface. In all Feynman diagrams, the
non-interacting single-particle propagators
$G_{\left(+\right)}^{(0)}$ and $G_{\left(-\right)}^{(0)}$ are
represented by a solid and a dashed line respectively, following
their association with the corresponding FS patches. They are given
by

\begin{eqnarray}
G_{(a)}^{(0)}(p)=\frac{\theta\left(\varepsilon_{a}(\mathbf{p})\right)}{p_{0}-\varepsilon_{a}(\mathbf{p})+i\delta}+
\frac{\theta\left(-\varepsilon_{a}(\mathbf{p})\right)}{p_{0}-\varepsilon_{a}(\mathbf{p})-i\delta}\label{gmais}
\end{eqnarray}

Before we deal directly with the one-particle irreducible function
$\Gamma_{R}^{(2,1)}$'s. In the next section we summarize the main
results of the field theory for the quasiparticle weight and the
coupling functions as well as the respective RG equations.

\section{Renormalized Coupling Functions and Self-energy up to two loops}

\subsection{Self-energy and quasiparticle weight up to two-loops}

\begin{figure}[b]
  \includegraphics[width=2.5in]{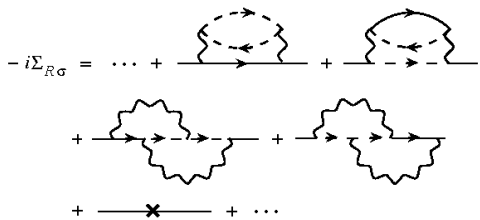}\\
  \caption{The diagrams for the self-energy up to two-loops.}\label{selfdia}
\end{figure}

In two-loop order another quantity plays an important role in the RG
equations for the renormalized couplings: the quasiparticle weight
$Z$. To calculate $Z$ we must determine the renormalized self-energy
$\Sigma_{R}$. Figure \ref{selfdia} displays the four important
$\Sigma_{R}$ diagrams for the determination of $Z$ up to two-loop
order. Those diagrams produce logarithmic singularities multiplied
by the factor $\left(p_{0}-v_{F}(|p_{\perp}|-k_{F})\right)$ which
can only be canceled out by the corresponding counterterm diagram
associated with the multiplicative fermion field factor Z. The use
of an appropriate RG prescription for the one-particles Green's
function at $p_{\perp}=\pm k_{F}$ together with those contributions
leads to the determination of $Z$. It then follows that the
quasiparticle weight satisfies the RG equation

\begin{equation}
\omega\frac{\partial
Z\left(p_{\parallel};\omega\right)}{\partial\omega}=\gamma
Z\left(p_{\parallel};\omega\right),\label{zrg}
\end{equation}

\noindent where $\gamma$ is the anomalous dimension. The full
expression for $\gamma$ is given in our Appendix A. The numerical
estimation of $Z$ follows our previous work. As we observe later the
suppression of the quasiparticle weight as we strengthen the
interactions changes dramatically the one-loop scenario.

\begin{figure}[t]
  \includegraphics[height=3.2in]{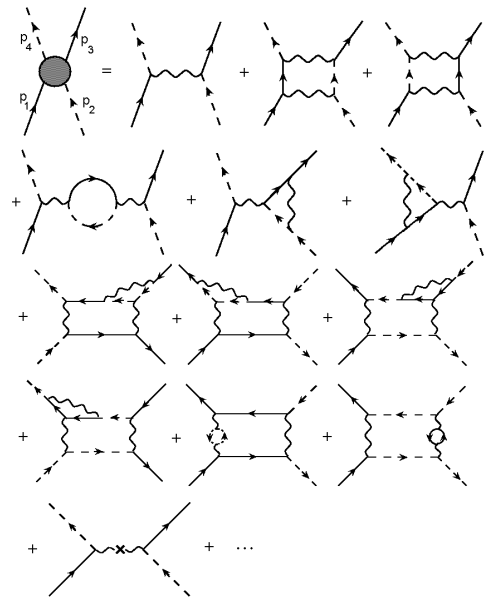}\\
  \caption{The diagrams for the renormalized four-point vertex in the backscattering channel.}\label{g1r}
\end{figure}

\subsection{Two-loop RG equations for the renormalized couplings}

\begin{figure}[t]
  \includegraphics[height=3.5in]{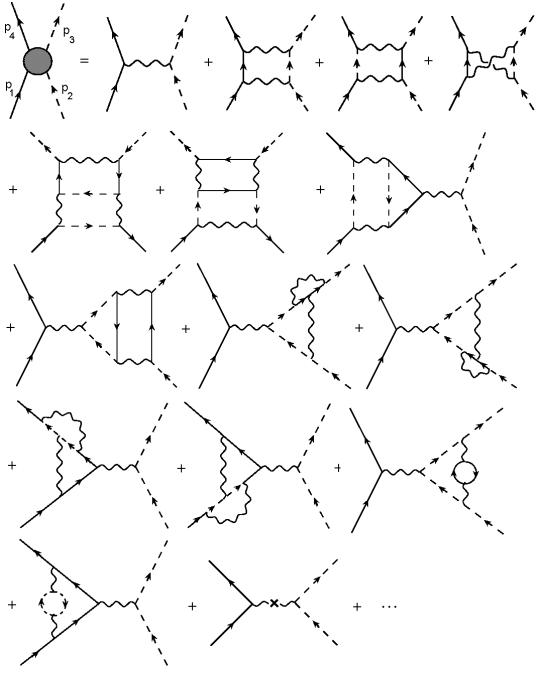}\\
  \caption{The diagrams for the renormalized four-point vertex in the forward scattering channel.}\label{g2r}
\end{figure}

In Eq. (\ref{lagr}) we wrote the renormalized Lagrangian which will
automatically generate renormalized physical quantities in
perturbation theory at any loop order. As a result, applying
suitable Feynman rules one can arrive at the diagrams shown in the
Fig. \ref{g1r} and Fig. \ref{g2r}. The last diagrams in both sets
are the counterterms which render the theory finite. In this way, we
identify the renormalized one-particle irreducible
$\Gamma_{iR}^{\left(4\right)}$ $(i=1,2)$ such that, at the FS, the
corresponding renormalized coupling functions
$g_{iR}\left(p_{1\parallel},p_{2\parallel},p_{3\parallel};\omega\right)$
are given by

\begin{equation}
\Gamma_{i}^{\left(4\right)}\left(p_{1},p_{2},p_{3}\right)|_{FS}=-ig_{iR}\left(p_{1},p_{2},p_{3};\omega\right),
\label{gamapresc}
\end{equation}

Taking into account the associated $Z$ factors for the external
momenta and the RG conditions $dg_{iB}/d\omega=0$ for the bare
coupling functions the RG equations for the coupling functions
follow immediately

\begin{eqnarray}
\omega\frac{dg_{iR}\left(p_{1\parallel},p_{2\parallel},p_{3\parallel}\right)}{d\omega}=&&\frac{1}{2}\sum_{j=1}^{4}\gamma\left(p_{j\parallel}\right)g_{iR}\left(p_{1\parallel},p_{2\parallel},p_{3\parallel}\right)
\nonumber \\&&-\omega\frac{\partial\Delta
g_{iR}\left(p_{1\parallel},p_{2\parallel},p_{3\parallel}\right)}{\partial\omega}.
\label{gir2l}
\end{eqnarray}

\noindent where $i=1,2$. We make use of the numerical estimates of
those RG equations in the calculations presented in this work.
Having said that we are now set to implement the RG strategy to
calculate the renormalized form factors which are indeed the linear
response functions of the system for the flat FS in 2D.

\section{Response Functions}

Following the RG strategy to study the pairing and density wave
instabilities of the system let us add initially to the renormalized
Lagrangian of the system two \emph{fictitious} infinitesimal
external fields $h_{SC}$(for the pairing term) and $h_{DW}$(for the
density wave) which act essentially as source fields for the
generation of particle-particle and particle-hole pairs. That is, we
include in our renormalized Lagrangian the contributions

\begin{align}
&L_{ext}=\frac{1}{V}\sum_{\mathbf{k},\mathbf{q}\atop
\alpha,\beta}\big[
Z^{1/2}(k_{\parallel})Z^{1/2}(q_{\parallel}-k_{\parallel})h_{SC}^{\alpha\beta}\left(\mathbf{q}\right)\mathcal{T}_{SC}^{B\alpha\beta}\left(\mathbf{k},\mathbf{q}\right)\nonumber \\
&\times\psi_{R(+)\alpha}^{\dagger}\left(\mathbf{k}\right)
\psi_{R(-)\beta}^{\dagger}\left(\mathbf{q-k}\right)
+Z^{1/2}(k_{\parallel})Z^{1/2}(k_{\parallel}-q_{\parallel})\nonumber
\\&\times h_{DW}^{\alpha\beta}\left(\mathbf{q}\right)\mathcal{T}_{DW}^{B\alpha\beta}\left(\mathbf{k},\mathbf{q}\right)\psi_{R(+)\alpha}^{\dagger}\left(\mathbf{k}\right)
\psi_{R(-)\beta}\left(\mathbf{k-q}\right)+H.c.\big].\nonumber\\\label{lagext}
\end{align}

\noindent where
$\mathcal{T}_{i}^{B\alpha\beta}\left(\mathbf{k},\mathbf{q}\right)$
$(i=SC,DW)$ is a \emph{bare} form factor which is determined in such
a way to incorporate the symmetry it is associated with. Notice that
in this model the external fields are made spin dependent since this
is important for the definition of some symmetries which we will
refer next. As we have done in section II we will consider again the
thermodynamic limit with all momenta summations becoming integrals.
By means of the added Lagrangian $L_{ext}$ we are now able to
generate the one-particle irreducible functions associated with the
composite pairing and the composite particle-hole operators. Since
we are interested in the response functions for the density wave and
superconductor channels we need to define the associated three
points generalized Green's functions namely
$G_{DW\alpha\beta}^{R(2,1)}$ and $G_{SC\alpha\beta}^{R(2,1)}$. In
doing this we get back the corresponding
$\Gamma_{DW\alpha\beta}^{R(2,1)}$ and
$\Gamma_{SC\alpha\beta}^{R(2,1)}$ by cutting out the external legs
of the corresponding $G_{i}^{R(2,1)}$'s. The $G_{i}^{R(2,1)}$'s are
given by

\begin{widetext}
\begin{equation}
G_{DW\alpha\beta}^{R(2,1)}\left(p,q\right)=i\frac{\delta}{\delta
h_{DW}^{\gamma\delta}\left(\mathbf{q}\right)}\left<\psi_{R(+)\alpha}^{\dagger}(\mathbf{p})\psi_{R(-)\beta}(\mathbf{p-q})exp\left[-i\int_{q_{0}p_{0}}L_{ext}\left[h_{DW}^{\gamma\delta},h_{SC}^{\gamma\delta}\right]\right]\right>_{h_{DW}^{\gamma\delta}=0}\label{dwc}
\end{equation}

\noindent and

\begin{equation}
G_{SC\alpha\beta}^{R(2,1)}\left(p,q\right)=i\frac{\delta}{\delta
h_{SC}^{\gamma\delta}\left(\mathbf{q}\right)}\left<\psi_{R(+)\alpha}^{\dagger}(\mathbf{p})\psi_{R(-)\beta}^{\dagger}(\mathbf{q-p})exp\left[-i\int_{q_{0}p_{0}}L_{ext}\left[h_{DW}^{\gamma\delta},h_{SC}^{\gamma\delta}\right]\right]\right>_{h_{SC}^{\gamma\delta}=0}\label{scc}
\end{equation}
\end{widetext}

\noindent where $<...>$ stands for
$\int\mathcal{D}\psi\mathcal{D}\psi^{\dagger}\exp
iS[\psi,\psi^{\dagger}]$( ... ) with S being the classical action
associated with the renormalized Lagrangian given by Eq.
(\ref{lagr}).

\begin{figure}[b]
  \includegraphics[height=3.0in]{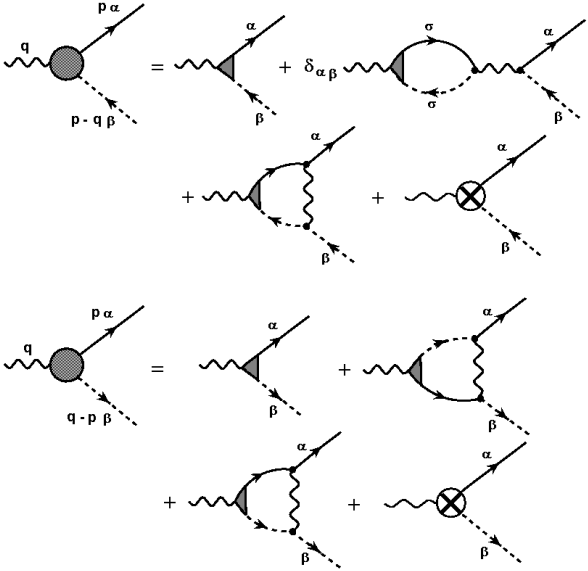}\\
  \caption{The Feynman diagrams up to one-loop order for the renormalized $\mathcal{T}_{DW}^{R\alpha\beta}$ in the density wave channel and the renormalized $\mathcal{T}_{SC}^{R\alpha\beta}$ in the superconductor channel.}\label{DWSC}
\end{figure}

Following conventional Feynman rules we display the diagrams in Fig.
\ref{DWSC} for the density wave and superconductor channels up to
one-loop order respectively.

Notice that the $\Gamma_{R}^{(2,1)}$ counterterm diagrams cancel
exactly all the corresponding higher order diagrams which scale as
$(\ln(\Omega/\omega))^{n}$ with $n\geq2$. Moreover, there are no
other linearly $\log$ divergent diagrams in $\Gamma_{R}^{(2,1)}$.
Consequently, in our scheme the resulting higher loop effects in
higher order contributions for $\Gamma_{R}^{(2,1)}$ are produced by
the $Z$ factors in our RG equations. Hence, if we consider the
external Lagrangian (\ref{lagext}) we can rewrite the \emph{bare}
form factors such that

\begin{subequations}
\begin{align}
\mathcal{T}_{DW}^{B\alpha\beta}\left(\mathbf{p},\mathbf{q}\right)&=Z^{-1/2}(p_{\parallel},\omega)Z^{-1/2}(p_{\parallel}-q_{\parallel},\omega)\nonumber \\
&\times Z_{DW}^{-1}(p_{\parallel},q_{\parallel},\omega)\mathcal{T}_{DW}^{R\alpha\beta}(\mathbf{p},\mathbf{q})\label{hdwa}\\
\mathcal{T}_{SC}^{B\alpha\beta}\left(\mathbf{p},\mathbf{q}\right)&=Z^{-1/2}(p_{\parallel},\omega)Z^{-1/2}(q_{\parallel}-p_{\parallel},\omega)\nonumber
\\&\times Z_{SC}^{-1}(p_{\parallel},q_{\parallel},\omega)\mathcal{T}_{SC}^{R\alpha\beta}(\mathbf{p},\mathbf{q})\label{hsca}
\end{align}
\end{subequations}

\noindent or, equivalently

\begin{subequations}
\begin{align}
\mathcal{T}_{DW}^{B\alpha\beta}\left(\mathbf{p},\mathbf{q}\right)&=Z^{-1/2}(p_{\parallel},\omega)Z^{-1/2}(p_{\parallel}-q_{\parallel},\omega)\nonumber \\ &\times\left[\mathcal{T}_{DW}^{R\alpha\beta}(\mathbf{p},\mathbf{q})+\Delta\mathcal{T}_{DW}^{R\alpha\beta}(\mathbf{p},\mathbf{q})\right]\label{hdw}\\
\mathcal{T}_{SC}^{B\alpha\beta}\left(\mathbf{p},\mathbf{q}\right)&=Z^{-1/2}(p_{\parallel},\omega)Z^{-1/2}(q_{\parallel}-p_{\parallel},\omega)\nonumber
\\&\times\left[\mathcal{T}_{SC}^{R\alpha\beta}(\mathbf{p},\mathbf{q})+\Delta\mathcal{T}_{SC}^{R\alpha\beta}(\mathbf{p},\mathbf{q})\right]\label{hsc}
\end{align}
\end{subequations}

The counterterm $(\Delta\mathcal{T}_{DW(SC)}^{R\alpha\beta})$
guarantees the cancelation of the divergent vertex functions
diagrams in one-loop order. This renders the theory finite at the
FS. Notice that the renormalized form factors depend on the momenta
$\mathbf{p}$ and $\mathbf{q}$. Now, we are ready to set up the
prescriptions for the renormalized one-particle irreducible Green's
functions $\Gamma_{Ri\alpha\beta}^{(2,1)}$'s, in terms of
experimentally observable physical quantities

\begin{subequations}
\begin{align}
\Gamma_{DW\alpha\beta}^{R(2,1)}(p_{\parallel}, p_{0}=\omega,
p_{\perp}= k_{F}; q_{\parallel}, &q_{\perp}=2k_{F})=\nonumber
\\&-i\mathcal{T}_{DW}^{R\alpha\beta}(p_{\parallel},
q_{\parallel})\label{dwcond}\\
\Gamma_{SC\alpha\beta}^{R(2,1)}(p_{\parallel},p_{0}=\omega,p_{\perp}=
k_{F};q_{\parallel},&q_{\perp}=0)=\nonumber
\\&-i\mathcal{T}_{SC}^{R\alpha\beta}(p_{\parallel},q_{\parallel}).\label{sccond}
\end{align}
\end{subequations}

As we already mentioned all renormalized quantities are RG energy
scale dependent. However, to avoid overloading notations we omit
this dependence from now on except when it is strictly necessary to
do so. In this way, considering the diagrams shown in Fig \ref{DWSC}
and making use of the RG conditions stated in Eqs. (\ref{dwcond})
and (\ref{sccond}) we arrive at

\begin{subequations}
\begin{align}
\Delta\mathcal{T}_{DW}^{R\alpha\beta}&=-\frac{1}{4\pi^{2}v_{F}}\int_{\mathcal{D}_{1}}dk_{\parallel}\bigg[\sum_{\sigma=\uparrow\downarrow}g_{1R}(k_{\parallel},p_{\parallel}-q_{\parallel},p_{\parallel})\nonumber
\\ &\times\mathcal{T}_{DW}^{R\sigma\sigma}(k_{\parallel},q_{\parallel})-g_{2R}(k_{\parallel},p_{\parallel}-q_{\parallel},p_{\parallel})\nonumber \\ &\times\mathcal{T}_{DW}^{R\alpha\beta}(k_{\parallel},q_{\parallel})\bigg]\ln\left(\frac{\omega}{\Omega}\right)\label{ddw}\\
\Delta\mathcal{T}_{SC}^{R\alpha\beta}&=-\frac{1}{4\pi^{2}v_{F}}\int_{\mathcal{D}_{2}}dk_{\parallel}\big[g_{2R}(k_{\parallel},q_{\parallel}-k_{\parallel},q_{\parallel}-p_{\parallel})\nonumber
\\ &-g_{1R}(k_{\parallel},q_{\parallel}-k_{\parallel},p_{\parallel})\big]\mathcal{T}_{SC}^{R\alpha\beta}(k_{\parallel},q_{\parallel})\ln\left(\frac{\omega}{\Omega}\right)\label{dsc}
\end{align}
\end{subequations}

Consistently with the renormalized density wave and pairing form
factors $\mathcal{T}_{DW}^{R\alpha\beta}$ and
$\mathcal{T}_{SC}^{R\alpha\beta}$ we can now do the symmetrization
with respect the spin components to define
\begin{subequations}
\begin{align}
&\mathcal{T}_{CDW}^{R}(p_{\parallel},q_{\parallel}) =
\mathcal{T}_{DW}^{R\uparrow\uparrow}(p_{\parallel},q_{\parallel}) +
\mathcal{T}_{DW}^{R\downarrow\downarrow}(p_{\parallel},q_{\parallel})\label{cdw}\\
&\mathcal{T}_{SDW}^{R}(p_{\parallel},q_{\parallel}) =
\mathcal{T}_{DW}^{R\uparrow\uparrow}(p_{\parallel},q_{\parallel}) -
\mathcal{T}_{DW}^{R\downarrow\downarrow}(p_{\parallel},q_{\parallel})\label{sdw}\\
&\mathcal{T}_{SSC}^{R}(p_{\parallel},q_{\parallel}) =
\mathcal{T}_{SC}^{R\uparrow\downarrow}(p_{\parallel},q_{\parallel})
-
\mathcal{T}_{SC}^{R\downarrow\uparrow}(p_{\parallel},q_{\parallel})\label{ssc}\\
&\mathcal{T}_{TSC}^{R}(p_{\parallel},q_{\parallel}) =
\mathcal{T}_{SC}^{R\uparrow\downarrow}(p_{\parallel},q_{\parallel})
+
\mathcal{T}_{SC}^{R\downarrow\uparrow}(p_{\parallel},q_{\parallel})\label{tsc}
\end{align}
\end{subequations}

\noindent where (CDW) relates to charge density wave, (SDW) to spin
density wave, (SSC) to singlet superconductivity and (TSC) to
triplet superconductivity respectively. Now, we can differentiate
the equations (\ref{hdw}) and (\ref{hsc}) with respect to $\omega$
to get

\begin{subequations}
\begin{align}
\omega\frac{d}{d\omega}\mathcal{T}_{DW}^{B\alpha\beta}&=\omega\frac{d}{d\omega}\bigg[\left(Z^{-1/2}(p_{\parallel};\omega)\right)\left(Z^{-1/2}(p_{\parallel}-q_{\parallel};\omega)\right)\nonumber \\
&\times\left(\mathcal{T}_{DW}^{R\alpha\beta}(p_{\parallel},q_{\parallel})+\Delta\mathcal{T}_{DW}^{R\alpha\beta}(p_{\parallel},q_{\parallel})\right)\bigg]\label{dhdw}\\
\omega\frac{d}{d\omega}\mathcal{T}_{SC}^{B\alpha\beta}&=\omega\frac{d}{d\omega}\bigg[\left(Z^{-1/2}(p_{\parallel};\omega)\right)\left(Z^{-1/2}(q_{\parallel}-p_{\parallel};\omega)\right)\nonumber \\
&\times\left(\mathcal{T}_{SC}^{R\alpha\beta}(p_{\parallel},q_{\parallel})+\Delta\mathcal{T}_{SC}^{R\alpha\beta}(p_{\parallel},q_{\parallel})\right)\bigg]\label{dhsc}
\end{align}
\end{subequations}

\noindent Taking into account the fact that the \emph{bare}
quantities do not know anything about the RG scale we can arrive at

\begin{subequations}
\begin{align}
&\omega\frac{d}{d\omega}\mathcal{T}_{DW}^{R}(p_{\parallel},q_{\parallel})=-\omega\frac{d}{d\omega}\Delta\mathcal{T}_{DW}^{R}(p_{\parallel},q_{\parallel})\nonumber
\\&+\frac{1}{2}\mathcal{T}_{DW}^{R}(p_{\parallel},q_{\parallel})\left[\gamma(p_{\parallel},\omega)+\gamma(p_{\parallel}-q_{\parallel},\omega)\right]\label{rgdw}\\
&\omega\frac{d}{d\omega}\mathcal{T}_{SC}^{R}(p_{\parallel},q_{\parallel})=-\omega\frac{d}{d\omega}\Delta\mathcal{T}_{SC}^{R}(p_{\parallel},q_{\parallel})\nonumber
\\&+\frac{1}{2}\mathcal{T}_{SC}^{R}(p_{\parallel},q_{\parallel})\left[\gamma(p_{\parallel},\omega)+\gamma(q_{\parallel}-p_{\parallel},\omega)\right]\label{rgsc}
\end{align}
\end{subequations}

\noindent where $\gamma$ is the anomalous dimension. Since we have
already defined the symmetrized renormalized form factors with
respect to spin projection, we are now able to write down their
corresponding RG equations

\begin{subequations}
\begin{align}
\omega\frac{d}{d\omega}\mathcal{T}_{b}^{R}(p_{\parallel},q_{\parallel})&=-\omega\frac{d}{d\omega}\Delta\mathcal{T}_{b}^{R}(p_{\parallel},q_{\parallel})+\frac{1}{2}\mathcal{T}_{b}^{R}(p_{\parallel},q_{\parallel})\nonumber
\\&\times\left[\gamma(p_{\parallel},\omega)+\gamma(p_{\parallel}-q_{\parallel},\omega)\right]\label{rga}\\
\omega\frac{d}{d\omega}\mathcal{T}_{c}^{R}(p_{\parallel},q_{\parallel})&=-\omega\frac{d}{d\omega}\Delta\mathcal{T}_{c}^{R}(p_{\parallel},q_{\parallel})+\frac{1}{2}\mathcal{T}_{c}^{R}(p_{\parallel},q_{\parallel})\nonumber
\\&\times\left[\gamma(p_{\parallel},\omega)+\gamma(q_{\parallel}-p_{\parallel},\omega)\right]\label{rgb}
\end{align}
\end{subequations}

\begin{figure}[b]
  \includegraphics[height=1.3in]{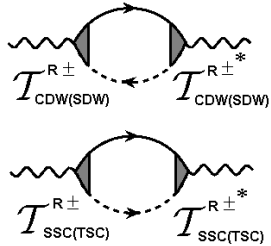}\\
  \caption{The respective susceptibilities generated from the renormalized form factors $\mathcal{T}_{CDW(SDW)}^{R\pm}$ and $\mathcal{T}_{SSC(TSC)}^{R\pm}$.}\label{suscet}
\end{figure}

\noindent where $b=CDW,SDW$ and $c=SSC,TSC$. The full expressions
for the symmetrized counterterms $\Delta \mathcal{T}_{i}^{R}$'s for
the anomalous dimension is given in Appendix A. Due to the
particular shape of our flat FS, the renormalized couplings must be
symmetrical with respect to the exchange of upper(right) and
lower(left) particles and the change of sign of the external
$p_{i\parallel}$'s. It then follows that

\begin{eqnarray}
g_{iR}(p_{1\parallel},p_{2\parallel},p_{3\parallel},p_{4\parallel})&=&g_{iR}(-p_{1\parallel},-p_{2\parallel},-p_{3\parallel},-p_{4\parallel})\nonumber
\\g_{iR}(p_{1\parallel},p_{2\parallel},p_{3\parallel},p_{4\parallel})&=&g_{iR}(p_{2\parallel},p_{1\parallel},p_{4\parallel},p_{3\parallel})\nonumber
\\g_{iR}(p_{1\parallel},p_{2\parallel},p_{3\parallel},p_{4\parallel})&=&g_{iR}(p_{4\parallel},p_{3\parallel},p_{2\parallel},p_{1\parallel})\label{sim1}
\end{eqnarray}

These conditions need to be satisfied by the RG equations in order
for them to produce the correct numerical results when we approach
the FS. These symmetries are satisfied by the RG equations
(\ref{rga}) and (\ref{rgb}) which are symmetrical with respect to
the sign reversal of $p_{\parallel}$ for a fixed $q_{\parallel}$.
Following this we can therefore define two irreducible
representations of this symmetry which never mix with each other

\begin{subequations}
\begin{eqnarray}
\mathcal{T}_{b}^{R\pm}(p_{\parallel},q_{\parallel})=\mathcal{T}_{b}^{R}(p_{\parallel},q_{\parallel})\pm\mathcal{T}_{b}^{R}(-p_{\parallel},q_{\parallel})\label{sim2}\\
\mathcal{T}_{c}^{R\pm}(p_{\parallel},q_{\parallel})=\mathcal{T}_{c}^{R}(p_{\parallel},q_{\parallel})\pm\mathcal{T}_{c}^{R}(-p_{\parallel},q_{\parallel})\label{sim3}
\end{eqnarray}
\end{subequations}

\noindent where again $b=CDW,SDW$ and $c=SSC,TSC$. The $(+)$ sign is
associated with the s-wave symmetry whereas the $(-)$ sign is
associated with the d-wave symmetry instead. Now, considering the
symmetries of the couplings (\ref{sim1}) and the new symmetrized
form factors (\ref{sim2}) and (\ref{sim3}) we can write the RG
equations (\ref{rga}) and (\ref{rgb}) in the following way

\begin{subequations}
\begin{align}
\omega\frac{d}{d\omega}\mathcal{T}_{b}^{R\pm}(p_{\parallel},q_{\parallel})&=-\omega\frac{d}{d\omega}\Delta\mathcal{T}_{b}^{R\pm}(p_{\parallel},q_{\parallel})+\frac{1}{2}\mathcal{T}_{b}^{R\pm}(p_{\parallel},q_{\parallel})\nonumber
\\&\times\left[\gamma(p_{\parallel},\omega)+\gamma(p_{\parallel}-q_{\parallel},\omega)\right]\label{rga1}\\
\omega\frac{d}{d\omega}\mathcal{T}_{c}^{R\pm}(p_{\parallel},q_{\parallel})&=-\omega\frac{d}{d\omega}\Delta\mathcal{T}_{c}^{R\pm}(p_{\parallel},q_{\parallel})+\frac{1}{2}\mathcal{T}_{c}^{R\pm}(p_{\parallel},q_{\parallel})\nonumber
\\&\times\left[\gamma(p_{\parallel},\omega)+\gamma(q_{\parallel}-p_{\parallel},\omega)\right]\label{rgb1}
\end{align}
\end{subequations}

\noindent where the expressions for
$\Delta\mathcal{T}_{CDW}^{R\pm}$, $\Delta\mathcal{T}_{SDW}^{R\pm}$,
$\Delta\mathcal{T}_{SSC}^{R\pm}$, $\Delta\mathcal{T}_{TSC}^{R\pm}$
can be found in the Appendix B. The plus sign in the DW's is
associated with the charge and spin density waves. However, the
minus sign symmetry of the parallel momentum along the FS in the
DW's yield circular charge(spin) currents flowing around the square
lattice with alternating directions. In this way we associate this
symmetry of the DW's with the charge and spin current waves also
known as flux phases. Once the renormalized
$\mathcal{T}_{CDW(SDW)}^{R\pm}$ and $\mathcal{T}_{SSC(TSC)}^{R\pm}$
are found we can define the related susceptibilities following the
diagrammatic scheme shown in Fig. \ref{suscet}. As one can see there
is a IR divergent \emph{bubble} in each channel. As a result we get

\begin{subequations}
\begin{align}
\chi_{b}^{R\pm}(q_{\parallel},\omega)=\frac{1}{4\pi^{2}v_{F}}\ln\left(\frac{\omega}{\Omega}\right)&\int_{\mathcal{D}_{3}}dp_{\parallel}\left(\mathcal{T}_{b}^{R\pm}(p_{\parallel},q_{\parallel})\right)^{*}\nonumber
\\ &\times\mathcal{T}_{b}^{R\pm}(p_{\parallel},q_{\parallel})\label{suscetib1}\\
\chi_{c}^{R\pm}(q_{\parallel},\omega)=\frac{1}{4\pi^{2}v_{F}}\ln\left(\frac{\omega}{\Omega}\right)&\int_{\mathcal{D}_{4}}dp_{\parallel}\left(\mathcal{T}_{c}^{R\pm}(p_{\parallel},q_{\parallel})\right)^{*}\nonumber
\\ &\times\mathcal{T}_{c}^{R\pm}(p_{\parallel},q_{\parallel})\label{suscetib2}
\end{align}
\end{subequations}

\noindent where $\mathcal{D}_{3}$ and $\mathcal{D}_{4}$ are the
intervals determined in the Appendix B and $b$ and $c$ refer to the
symmetries mentioned before. Notice that if we don't go beyond
one-loop order all RG equations for the susceptibilities approach
the strong coupling regime. As a result all susceptibilities diverge
in the IR limit at that order of perturbation theory. In two-loop
order we show that only one susceptibility at a time remains
divergent and all the others approach fixed \emph{plateaus} values
instead. Having said that we also call attention to the fact that in
two-loops the leading susceptibility approaches the strong coupling
regime in a much slower rate than the one found in one-loop order.
This might mean that the IR divergence may disappear altogether in
the presence of higher order corrections and as a result one should
not expect true long-range order if all this is taken into account.
We discuss this further later on. From Eqs. (\ref{suscetib1}) and
(\ref{suscetib2}) we arrive immediately at the RG equations for the
various susceptibilities

\begin{subequations}
\begin{align}
\omega\frac{d}{d\omega}\chi_{b}^{R\pm}(q_{\parallel},\omega)=\frac{1}{4\pi^{2}v_{F}}\int_{\mathcal{D}_{3}}&dp_{\parallel}\left(\mathcal{T}_{b}^{R\pm}(p_{\parallel},q_{\parallel})\right)^{*} \times\nonumber\\
&\mathcal{T}_{b}^{R\pm}(p_{\parallel},q_{\parallel})\label{suscetibr1}\\
\omega\frac{d}{d\omega}\chi_{c}^{R\pm}(q_{\parallel},\omega)=\frac{1}{4\pi^{2}v_{F}}\int_{\mathcal{D}_{4}}&dp_{\parallel}\left(\mathcal{T}_{c}^{R\pm}(p_{\parallel},q_{\parallel})\right)^{*}\times\nonumber\\
&\mathcal{T}_{c}^{R\pm}(p_{\parallel},q_{\parallel})\label{suscetibr2}
\end{align}
\end{subequations}

These expressions are written in general form. However, since we are
just considering $g_{1R}$ and $g_{2R}$ couplings it turns out that
$\mathcal{T}_{b,c}^{R\pm}=\left(\mathcal{T}_{b,c}^{R\pm}\right)^{*}$.

\section{Results}

\begin{figure}[b]
  \includegraphics[width=2.7in]{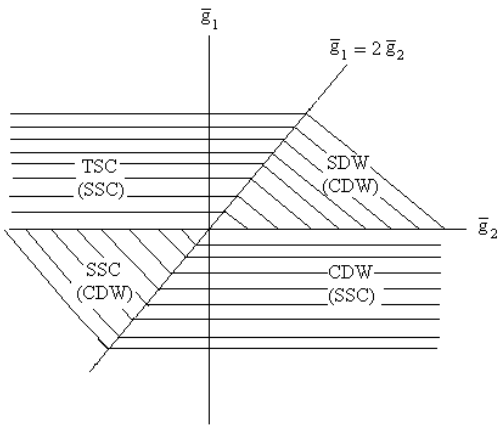}\\
  \caption{Phase diagram for the susceptibilities in one-dimension up to two-loops.}\label{dfase1}
\end{figure}

In order to solve all those RG equations we have to resort to
numerical methods since we want to estimate how the order parameters
change as we vary the scale $\omega$ to take the physical system
towards the FS. As we have done in an earlier work we discretize the
FS continuum replacing the interval $-\Delta \leqslant p_{\parallel}
\leqslant \Delta$ by a discrete set of 33 points. For convenience,
we use $\omega = \Omega\exp(-l)$, where $\Omega = 2k_{F}\lambda$
with $l$ being our RG step. We choose $\Omega/v_{F}\Delta=1$. In
view of our choice of points for the FS, we are only allowed to go
up to $l\approx2.8$ in the RG flow to avoid the distance $\omega$ to
the FS being smaller than the distance between neighboring points
since the discretization procedure no longer applies in this case.

\begin{figure}[t]
  \includegraphics[width=3.3in]{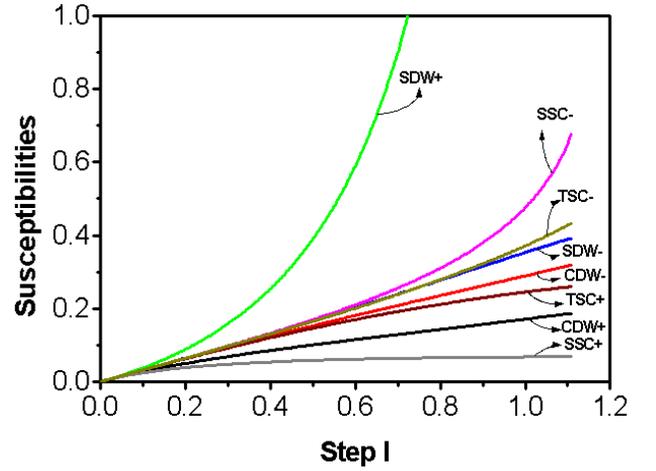}\\
  \caption{Susceptibilities $\chi_{a,b}^{\pm}(q_{\parallel}=0;l)$ against step l for one-loop approach with $\overline{g}_{1R}=\overline{g}_{2R}=10$ as initial conditions for the couplings. The SDW s-type diverges for l=1.2.}\label{susc1l}
\end{figure}

\begin{figure}[b]
  \includegraphics[width=3.3in]{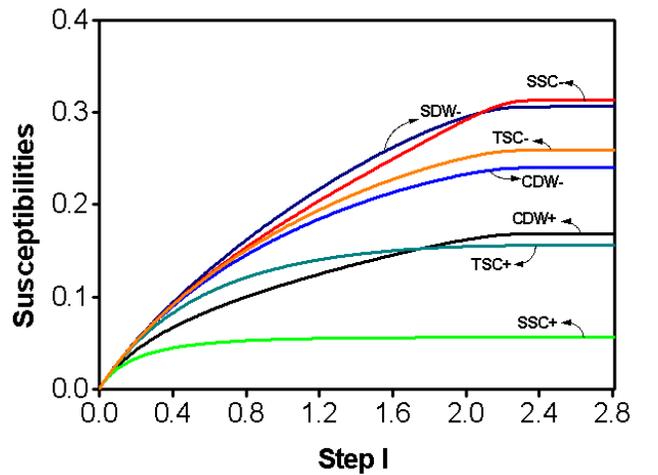}\\
  \caption{Susceptibilities $\chi_{a,b}^{\pm}(q_{\parallel}=0;l)$ against step l for two-loop approach with $\overline{g}_{1R}=\overline{g}_{2R}=10$ as initial conditions for the couplings.}\label{susc2l}
\end{figure}

The RG equations for all couplings considered here were reproduced
in section III. As a result all RG equations (\ref{zrg}),
(\ref{gir2l}), (\ref{rga}) and (\ref{rgb}) and (\ref{suscetibr1})
and (\ref{suscetibr2}) have to be solved simultaneously. Hence, the
numerical estimates become much more involved.

To improve the understanding of our results we also consider the
one-dimensional limits of all our RG equations. We reproduce all
those one-dimensional RG equations in our appendix C. Following the
same scheme used for $D=2$ all one-dimensional RG equations are
solved numerically in a self-consistent manner. Notice that as
expected the equation $\omega d(g_{1R}-2g_{2R})/d\omega=0$ is
trivially satisfied. In Fig. \ref{dfase1} we display the
one-dimensional phase diagram in coupling space. This phase diagram
is constructed in the usual way taking into account the two dominant
susceptibilities that diverge for a given initial values of the
renormalized couplings $g_{1R}$ and $g_{2R}$. We do that because
there are always two divergent susceptibilities in all region of the
phase space. Our numerical estimates are in good agreement with
Solyom's earlier RG results\cite{Solyom}. However, we noticed that
those divergences take place only when the RG step is too large or,
equivalently, $\omega$ is very small. This effect is amplified as we
move from one-loop to two-loops. As a result despite the fact that
there might be strong collective fluctuations in such a system we
cannot say that they are of \emph{long range} type since when we
consider higher orders in our perturbation theory the divergences
only take place in even larger RG steps. This situation will reflect
itself more emphatically in the two-dimensional case as we will see
next.

\begin{figure}[b]
  \includegraphics[width=3.3in]{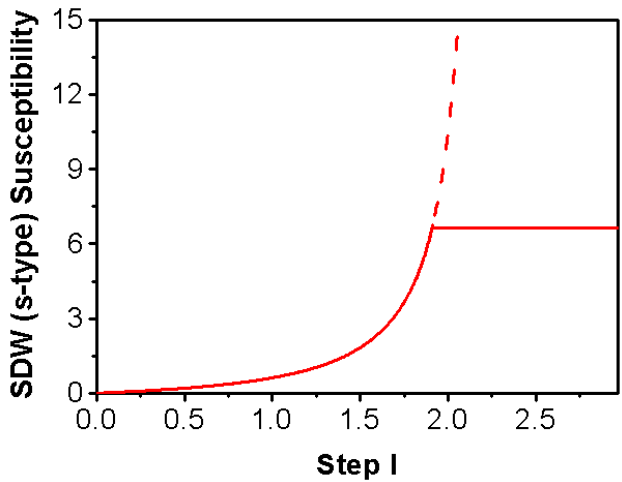}\\
  \caption{SDW+ susceptibility $\chi_{SDW}^{+}(q_{\parallel};l)$ with $\overline{g}_{1R}=\overline{g}_{2R}=10$ as initial conditions for the couplings. The SDW s-type diverges for $l=2.15$. Dashed line refers to perfect nesting of the FS. Solid line refers to the RG step $l=l_{cutoff}$ considering a ``corrugation'' in the FS.}\label{sdw2l}
\end{figure}

Now, we move on to our original RG equations in two-dimensions. The
choice of the initial conditions at $l=0$ in the RG equations are,
in principle, arbitrary. However, as we have done before, we set the
couplings equal to each other to mimic the Hubbard on-site repulsive
interaction parameter U in our RG scheme. Namely
$\overline{g}_{1R}=\overline{g}_{2R}=10$, where
$\overline{g}_{iR}=g_{iR}/\pi v_{F}$. In another paper we give more
details about this connection with the Hubbard model\cite{isl}.

In order to reproduce some symmetries of the order parameters with
respect to FS we choose the following initial conditions $(l=0)$

\begin{subequations}
\begin{align}
\mathcal{T}_{CDW}^{R+}(p_{\parallel},q_{\parallel})=&\mathcal{T}_{SDW}^{R+}(p_{\parallel},q_{\parallel})=\mathcal{T}_{SSC}^{R+}(p_{\parallel},q_{\parallel})=\nonumber
\\&\mathcal{T}_{TSC}^{R+}(p_{\parallel},q_{\parallel})=1\label{initc1}\\
\mathcal{T}_{CDW}^{R-}(p_{\parallel},q_{\parallel})=&\mathcal{T}_{SDW}^{R-}(p_{\parallel},q_{\parallel})=\mathcal{T}_{SSC}^{R-}(p_{\parallel},q_{\parallel})=\nonumber
\\&\mathcal{T}_{TSC}^{R-}(p_{\parallel},q_{\parallel})=\sqrt{2}\sin\left(\frac{\pi p_{\parallel}}{2}\right)\label{initc2}
\end{align}
\end{subequations}

Notice that despite the fact that the initial values of the form
factor are either unity or a function of $p_{\parallel}$ the
$q_{\parallel}$ dependence is generated naturally by the
renormaliztion process. The choice made for the symmetrized $(+)$
vertices is motivated by their independence with respect to the
change of sign of $p_{\parallel}$. In contrast with the
antisymmetric vertices our choice is oriented by our need to
reproduce the symmetries of the flux phases and the
$d_{x^{2}-y^{2}}$ superconductivity with respect to the FS.
Furthermore, we take all susceptibilities equal to zero at $l=0$,
that is, $\chi_{i}^{\pm}(q_{\parallel};l=0)=0$ $(i=CDW, SDW, SSC,
TSC)$.

\begin{figure}[b]
  \includegraphics[width=3.3in]{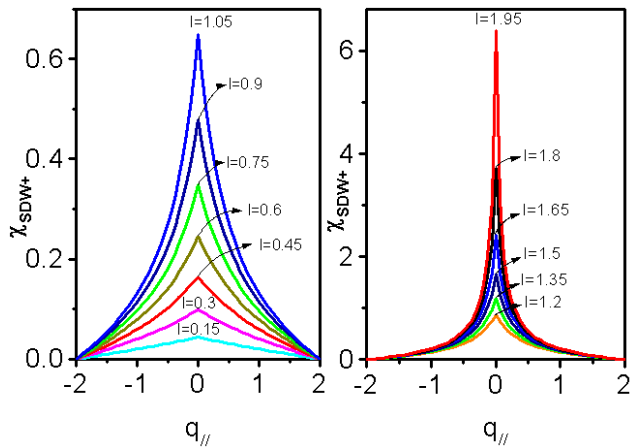}\\
  \caption{SDW+ susceptibility $\chi_{SDW}^{+}(q_{\parallel};l)$ against $q_{\parallel}$  for several RG steps with $\overline{g}_{1R}=\overline{g}_{2R}=10$ as initial conditions for the couplings.}\label{sdw2l2}
\end{figure}

To arrive at the final results we do a simultaneous calculation of
the renormalized form factors and the corresponding susceptibilities
together with the flow equations for the coupling functions $g_{1R}$
and $g_{2R}$. It emerges from our numerical estimates that the SDW
of symmetry type s produces the dominant susceptibility when the
transfer momentum is equal to the nesting vector
$(\mathbf{q}=\mathbf{Q}^{*})$. To make the comparison of our results
with previous estimates found in the literature we display our
computations in two steps. Initially we show the one-loop order
results since they can be compared directly with calculations
presented by other groups. Next, we move on to display our two-loop
calculations. Following that we discuss the most interesting results
and the difference between our estimates in one-loop and two-loops.

\subsection{One-loop RG approach}

The one-loop results for all susceptibilities are displayed in Fig.
\ref{susc1l}. As can be seen all susceptibilities diverge but the
leading one is the $SDW+$ symmetry in the repulsive HM like regime.
According to this, we should expect an insulating spin density wave
state and no sign of even nonconventional metallic behavior in the
physical system. It is interesting to observe that the second most
pronounced renormalized susceptibility corresponds to a
$d_{x^{2}-y^{2}}$ superconductivity symmetry.

Our result is consistent with other approaches based on one-loop
fermionic functional RG\cite{Metzner} and the so-called parquet
method\cite{Dzyaloshinskii}. In all those approaches, the feedback
of the quasiparticle weight $Z$ is not taken into account. The fact
that $Z\rightarrow0$ produces drastic changes in the flow of all RG
equations in two-loops as we will see next.

\begin{figure}[t]
  \includegraphics[width=3.0in]{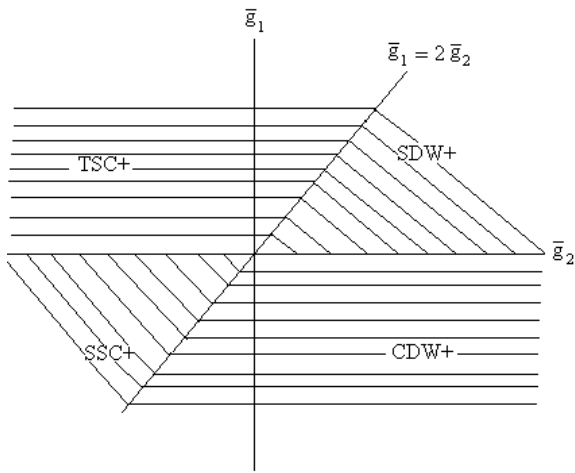}\\
  \caption{The leading susceptibilities for several initial values of coupling functions.}\label{diagfase}
\end{figure}

\subsection{The full two-loop approach}

In this final case, we solve the Eqs. (\ref{rga}) and (\ref{rgb})
and Eqs. (\ref{suscetibr1}) and (\ref{suscetibr2}) simultaneously
with the RG equations for the renormalized couplings using the same
initial conditions given by Eqs. (\ref{initc1}) and (\ref{initc2}),
namely, $\chi_{i}^{\pm}(q_{\parallel};l=0)=0$ and
$\overline{g}_{1R}=\overline{g}_{2R}=10$ respectively.

The results obtained for all susceptibilities excluding the
$\chi_{SDW+}$ are shown in Fig. \ref{susc2l}. As one can see the
effect of $Z$ in the RG equations changes drastically the one-loop
scenario. Notice that the \emph{plateaus} values appear at $l=2.2$
for almost all susceptibilities, that is, when the strong
renormalization of the coupling functions takes place. In contrast
with the picture described for the coupling functions at the same
two-loop order, these \emph{plateaus} are real fixed points since
their values are not sensitive to our FS discretization procedure.
Thus, we can infer from this, that with the exception of the $SDW+$
symmetry, all existing order parameters appear as \emph{short range
correlations}. However, for this HM like scenario, the $SDW+$
symmetry manifests itself in a completely different manner and to
illustrate this we plotted the $SDW+$ susceptibility as a function
of the parallel transfer momentum $(q_{\parallel})$ in Fig.
\ref{sdw2l2}.

As $q_{\parallel}$ approaches zero the $SDW+$ susceptibility
increases drastically as a function of the RG step $l$. In contrast
for any other value for the parallel transfer momentum
$(q_{\parallel}\neq0)$ the RG equation for $\chi_{SDW}^{R+}$ flows
to a given fixed value. Does this means that we can assign to the
$SDW+$ divergence at $q_{\parallel}=0$ a \emph{long-range} type
correlation? As we emphasized earlier the effect of the
quasiparticle weight in two-loop approach the $SDW+$ divergence
becomes predominant for larger values of the renormalization step
$l$. This is a strong evidence that even this divergence might not
be what it seems. In fact, we expect that if we were to consider
higher order contributions in perturbation theory this divergence
would occur for even larger $l$ values. In order to interpret
correctly the meaning of this divergent susceptibility we have to
keep in mind the limitations of our RG analysis. From the start we
said that we were dealing with a two-dimensional flat FS with
rounded corners. This FS represents a doped electron system slightly
away from half-filling. As a result there exists a non-zero chemical
potential $\mu$ which is an indirect measure of that doping regime.
$\mu$ introduces a new scale into the problem. At RG scales
$\omega>>\mu$ we can safely treat the FS as being made of perfectly
flat patches with rounded corners. As $\omega$ is lowered (i.e. as
we increase the number of RG l steps) our RG resolution of FS must
be corrected to include its ``corrugation'' effects. To take this
resolution effect approximately into account we follow Zheleznyak
\emph{et al}\cite{Dzyaloshinskii} in establishing a cutoff $l_{c}$
beyond which we are no longer able to be sure about the perfect
flatness of our FS.

We take $\mu$ such that $l_{c}=\ln(\Omega_{cutoff}/|\mu|)=1.95$
since $\chi_{SDW}^{R+}$ only diverges for $l=2.15$. This means that
for $l=l_{c}$ the SDW s-type will now reach a \emph{plateau} value
before this divergence takes place. We display this new
$\chi_{SDW}^{R+}$ in Figure \ref{sdw2l}. This excludes the existence
of long-range order in such a lightly doped Hubbard model like
regime.

In Figure \ref{diagfase} we display a phase diagram with the leading
susceptibilities for several initial conditions of the coupling
functions. In our calculations we notice that when we set initial
values for the couplings nearby each axis the symmetry associated
with the next phase becomes more pronounced. As an example, if we
are in the $SDW+$ phase and we choose initial conditions nearby the
straight line $\overline{g}_{1}=2\overline{g}_{2}$ we obtain as a
result a increase in the $TSC+$ susceptibility and so on. Moreover,
we can emphasize that this phase diagram is quite different from
that in one-dimension. The latter shows the possible ground states
inferred from the renormalized form factors which in turn are a
mixture of two states while in two-dimensions there is only one
pronounced susceptibility at a time which we associate with a strong
collective oscillation. This shows how much the one-dimension
scenario can change as one goes from one to two-dimensions.

\section{Conclusion}

In this paper we have performed a full two-loop theoretical RG
calculation for the symmetries $CDW\pm,SDW\pm,SSC\pm$ and $TSC\pm$
in the presence of a square two dimensional FS with rounded corners.
We neglected Umklapp effects at this stage since we are not at
half-filling. To render the theory finite we applied the RG field
theory by adding appropriate counterterms order by order in
perturbation theory.

Due to the particular shape of our FS we can reproduce some
symmetries by just choosing appropriate initial conditions for the
renormalized form factors. The general expectation is that the
resulting physical state has several competing instabilities such as
the flux phases ($CDW-$ and $SDW-$), the singlet $d_{x^{2}-y{^2}}$
superconductivity $(SSC-)$, the d-type triplet superconductivity
$(TSC-)$, the singlet and triplet superconductivity of s-type
($SSC+$ and $TSC+$) and finally the spin density wave $(SDW+)$ and
charge density wave $(CDW+)$. At one-loop order all those symmetries
diverge and the leading divergence is $SDW+$ suggesting that there
is a \emph{long-range} type correlation.

In two-loop order for the repulsive HM like initial conditions we
obtained that all susceptibilities go to stable fixed points except
the $SDW+$ symmetry that continues to diverge in this case. In order
to interpret correctly the meaning of this divergent susceptibility
we call attention to the fact that the divergence in two-loop order
takes place for larger values of the RG step $l$ in comparison with
what is obtained in one-loop order. It is therefore reasonable to
expect that in higher orders the divergence takes place for even
higher values of $l$. This happens due to the limitation of the RG
analysis.

In this model we assumed that the FS is completely flat and, as a
result, perfectly nested. However, in real situations, this naive
assumption breaks down since most FS's are expected to exhibit a
finite curvature at some critical scale $\omega_{c}$ as one
approaches the low-energy limit. In such a condition the nesting is
destroyed. If this happens, the RG flow of the susceptibility
associated with the $SDW+$ order parameter will become interrupted
due to the fact that this quantity strongly relies on the nesting
properties of our FS model. Consequently, this result asserts that
there should be only \emph{short-range}  order in the system in
agreement with the general expectation that strong quantum
fluctuations should be a dominant feature in such low dimensional
systems.

We sketched the phase diagram for different renormalized coupling
constants initial conditions in two-dimensions. There is only one
pronounced susceptibility at a time which we associate with a
dominant collective oscillation. In order to test our method in well
known grounds we solved all RG equations for one-dimensional case.
We reproduce correctly the phase diagram displayed earlier by
Solyom\cite{Solyom} together with the possible ground states of the
system inferred by renormalized form factors.

To have a better idea about the nature of the dominant instability
we have to calculate the corresponding uniform susceptibilities
which are calculated taking the transfer momentum equal to zero
$(\mathbf{q}=(0,0))$. This has been done in a recent paper for the
particular case of the repulsive HM like initial conditions . We
obtained in case that the charge compressibility flows to zero, that
is, $\partial n/\partial \mu\rightarrow0$ and, hence, we could
indeed consider that the $SDW+$ divergence would lead to a symmetry
breaking contradicting all arguments presented previously. However,
another quantity flows to zero as well. When we calculate the
uniform spin susceptibility we find the opening of a spin gap as
$\chi_{spin}^{uniform}$ flows to zero. Hence, it is our belief that
the quantum regime associated with such a state is characteristic of
an insulating spin liquid. This conclusion is in agreement with the
results presented in this work.

This work was partially supported by the Conselho Nacional de
Desenvolvimento Científico e Tecnológico (CNPq).

\appendix

\section{}

In this appendix, we write down the explicit form of the $\Delta
\mathcal{T}_{i}^{R}$'s $(i=CDW,SDW,SSC,TSC)$ and $\gamma$'s which
are taken into account in the Eqs. (\ref{rga}), (\ref{rgb}),
(\ref{rga1}) and (\ref{rgb1}). We also give the several intervals of
integration that are considered throughout this work. They are
following
\begin{center}%
\begin{displaymath}
\mathcal{D}_{1}=\left\{%
\begin{array}{ll}
    -\Delta\leqslant k_{\parallel} \leqslant \Delta, & \hbox{} \\
    -\Delta\leqslant p_{\parallel} \leqslant \Delta, & \hbox{} \\
    -2\Delta\leqslant q_{\parallel} \leqslant 2\Delta, & \hbox{} \\
    -\Delta\leqslant p_{\parallel}-q_{\parallel} \leqslant \Delta. & \hbox{} \\
\end{array}%
\right.
\end{displaymath}
\begin{displaymath}
\mathcal{D}_{2}=\left\{%
\begin{array}{ll}
    -\Delta\leqslant k_{\parallel} \leqslant \Delta, & \hbox{} \\
    -\Delta\leqslant p_{\parallel} \leqslant \Delta, & \hbox{} \\
    -2\Delta\leqslant q_{\parallel} \leqslant 2\Delta, & \hbox{} \\
    -\Delta\leqslant q_{\parallel}-k_{\parallel} \leqslant \Delta, & \hbox{} \\
    -\Delta\leqslant q_{\parallel}-p_{\parallel} \leqslant \Delta. & \hbox{} \\
\end{array}%
\right.
\end{displaymath}
\begin{displaymath}
\mathcal{D}_{3}=\left\{%
\begin{array}{ll}
    -\Delta\leqslant p_{\parallel} \leqslant \Delta, & \hbox{} \\
    -2\Delta\leqslant q_{\parallel} \leqslant 2\Delta, & \hbox{} \\
    -\Delta\leqslant p_{\parallel}-q_{\parallel} \leqslant \Delta. & \hbox{} \\
\end{array}%
\right.
\end{displaymath}
\begin{displaymath}
\mathcal{D}_{4}=\left\{%
\begin{array}{ll}
    -\Delta\leqslant p_{\parallel} \leqslant \Delta, & \hbox{} \\
    -2\Delta\leqslant q_{\parallel} \leqslant 2\Delta, & \hbox{} \\
    -\Delta\leqslant q_{\parallel}-p_{\parallel} \leqslant \Delta. & \hbox{} \\
\end{array}%
\right.
\end{displaymath}
\begin{displaymath}
\mathcal{D}_{5}=\left\{%
\begin{array}{ll}
    -\Delta\leqslant k_{\parallel} \leqslant \Delta, & \hbox{} \\
    -\Delta\leqslant p_{\parallel} \leqslant \Delta, & \hbox{} \\
    -\Delta\leqslant q_{1\parallel} \leqslant \Delta, & \hbox{} \\
    -\Delta\leqslant -k_{\parallel}+p_{\parallel}+q_{1\parallel} \leqslant \Delta. & \hbox{} \\
\end{array}%
\right.
\end{displaymath}
\begin{displaymath}
\mathcal{D}_{6}=\left\{%
\begin{array}{ll}
    -\Delta\leqslant k_{\parallel} \leqslant \Delta, & \hbox{} \\
    -\Delta\leqslant p_{\parallel} \leqslant \Delta, & \hbox{} \\
    -2\Delta\leqslant q_{\parallel} \leqslant 2\Delta, & \hbox{} \\
    -\Delta\leqslant q_{1\parallel} \leqslant \Delta, & \hbox{} \\
    -\Delta\leqslant -k_{\parallel}+p_{\parallel}-q_{\parallel}+q_{1\parallel} \leqslant \Delta. & \hbox{} \\
\end{array}%
\right.
\end{displaymath}
\begin{displaymath}
\mathcal{D}_{7}=\left\{%
\begin{array}{ll}
    -\Delta\leqslant k_{\parallel} \leqslant \Delta, & \hbox{} \\
    -2\Delta\leqslant q_{\parallel} \leqslant 2\Delta, & \hbox{} \\
    -\Delta\leqslant p_{\parallel} \leqslant \Delta, & \hbox{} \\
    -\Delta\leqslant q_{1\parallel} \leqslant \Delta, & \hbox{} \\
    -\Delta\leqslant -k_{\parallel}+q_{\parallel}-p_{\parallel}+q_{1\parallel} \leqslant \Delta. & \hbox{} \\
\end{array}%
\right.
\end{displaymath}
\end{center}

We begin with the expressions for the  $\Delta
\mathcal{T}_{i}^{R}$'s associated with $\Gamma^{(2,1)}$'s. We get

\begin{align}
&\Delta
\mathcal{T}_{CDW}^{R}(p_{\parallel},q_{\parallel})=\frac{1}{4\pi^{2}v_{F}}\ln\left(\frac{\Omega}{\omega}\right)\nonumber
\\ &\times \int_{\mathcal{D}_{1}} dk_{\parallel}\big[2g_{1R}
\left(k_{\parallel},p_{\parallel}-q_{\parallel},p_{\parallel}\right)\nonumber
\\&-g_{2R}\left(k_{\parallel},p_{\parallel}-q_{\parallel},k_{\parallel}-q_{\parallel}\right)\big]\mathcal{T}_{CDW}^{R}(k_{\parallel},q_{\parallel}),\label{delta1}
\end{align}

\begin{align}
&\Delta
\mathcal{T}_{SDW}^{R}(p_{\parallel},q_{\parallel})=-\frac{1}{4\pi^{2}v_{F}}\ln\left(\frac{\Omega}{\omega}\right)\nonumber
\\ &\times \int_{\mathcal{D}_{1}} dk_{\parallel}g_{2R}\left(k_{\parallel},p_{\parallel}-q_{\parallel},k_{\parallel}-q_{\parallel}\right)\mathcal{T}_{SDW}^{R}(k_{\parallel},q_{\parallel}),\label{delta2}
\end{align}

\begin{align}
&\Delta
\mathcal{T}_{SSC}^{R}(p_{\parallel},q_{\parallel})=\frac{1}{4\pi^{2}v_{F}}\ln\left(\frac{\Omega}{\omega}\right)\nonumber
\\ &\times \int_{\mathcal{D}_{2}} dk_{\parallel}\big[g_{1R}
\left(k_{\parallel},-k_{\parallel}+q_{\parallel},p_{\parallel}\right)\nonumber
\\&+g_{2R}\left(k_{\parallel},-k_{\parallel}+q_{\parallel},-p_{\parallel}+q_{\parallel}\right)\big]\mathcal{T}_{SSC}^{R}(k_{\parallel},q_{\parallel}),\label{delta3}
\end{align}

\noindent and finally

\begin{align}
&\Delta
\mathcal{T}_{TSC}^{R}(p_{\parallel},q_{\parallel})=-\frac{1}{4\pi^{2}v_{F}}\ln\left(\frac{\Omega}{\omega}\right)\nonumber
\\ &\times \int_{\mathcal{D}_{2}} dk_{\parallel}\big[g_{1R}
\left(k_{\parallel},-k_{\parallel}+q_{\parallel},p_{\parallel}\right)\nonumber
\\&-g_{2R}\left(k_{\parallel},-k_{\parallel}+q_{\parallel},-p_{\parallel}+q_{\parallel}\right)\big]\mathcal{T}_{TSC}^{R}(k_{\parallel},q_{\parallel}).\label{delta4}
\end{align}

The anomalous dimension $\gamma$ used in our RG equations for the
renormalized form factors is given by\cite{nosso}

\begin{align}
&\gamma(p_{\parallel};\omega)=\frac{1}{32\pi^{4}v_{F}^{2}}
\int_{\mathcal{D}_{5}}dk_{\parallel}
dq_{1\parallel}[2g_{1R}(-k_{\parallel}+p_{\parallel}+q_{1\parallel},\nonumber
\\&k_{\parallel},q_{1\parallel})g_{1R}\left(p_{\parallel},q_{1\parallel},k_{\parallel}\right)
+2g_{2R}\left(p_{\parallel},q_{1\parallel},-k_{\parallel}+p_{\parallel}+q_{1\parallel}\right)\nonumber
\\&\times g_{2R}\left(k_{\parallel},-k_{\parallel}+p_{\parallel}+q_{1\parallel},q_{1\parallel}\right)
-g_{1R}\left(p_{\parallel},q_{1\parallel},k_{\parallel}\right)\nonumber
\\ &\times g_{2R}\left(k_{\parallel},-k_{\parallel}+p_{\parallel}+q_{1\parallel},q_{1\parallel}\right)
-g_{2R}(p_{\parallel},q_{1\parallel},\nonumber
\\&-k_{\parallel}+p_{\parallel}+q_{1\parallel})g_{1R}\left(k_{\parallel},-k_{\parallel}+p_{\parallel}+q_{1\parallel},p_{\parallel}\right)],\label{gama1}
\end{align}

\section{}

Considering the symmetries obeyed by the coupling functions we get
the symmetrized $(\pm)$ renormalized form factors whose counterterms
$\Delta \mathcal{T}_{i}^{R\pm}$ are given by

\begin{align}
&\Delta
\mathcal{T}_{CDW}^{R\pm}(p_{\parallel},q_{\parallel})=\frac{1}{4\pi^{2}v_{F}}\ln\left(\frac{\Omega}{\omega}\right)\nonumber
\\ &\times \int_{\mathcal{D}_{1}} dk_{\parallel}\big[2g_{1R}
\left(k_{\parallel},p_{\parallel}-q_{\parallel},p_{\parallel}\right)\nonumber
\\&-g_{2R}\left(k_{\parallel},p_{\parallel}-q_{\parallel},k_{\parallel}-q_{\parallel}\right)\big]\mathcal{T}_{CDW}^{R\pm}(k_{\parallel},q_{\parallel}),\label{delta1}
\end{align}

\begin{align}
&\Delta
\mathcal{T}_{SDW}^{R\pm}(p_{\parallel},q_{\parallel})=-\frac{1}{4\pi^{2}v_{F}}\ln\left(\frac{\Omega}{\omega}\right)\nonumber
\\ &\times \int_{\mathcal{D}_{1}} dk_{\parallel}g_{2R}\left(k_{\parallel},p_{\parallel}-q_{\parallel},k_{\parallel}-q_{\parallel}\right)\mathcal{T}_{SDW}^{R\pm}(k_{\parallel},q_{\parallel}),\label{delta2}
\end{align}

\begin{align}
&\Delta
\mathcal{T}_{SSC}^{R\pm}(p_{\parallel},q_{\parallel})=\frac{1}{4\pi^{2}v_{F}}\ln\left(\frac{\Omega}{\omega}\right)\nonumber
\\ &\times \int_{\mathcal{D}_{2}} dk_{\parallel}\big[g_{1R}
\left(k_{\parallel},-k_{\parallel}+q_{\parallel},p_{\parallel}\right)\nonumber
\\&+g_{2R}\left(k_{\parallel},-k_{\parallel}+q_{\parallel},-p_{\parallel}+q_{\parallel}\right)\big]\mathcal{T}_{SSC}^{R\pm}(k_{\parallel},q_{\parallel}),\label{delta3}
\end{align}

\noindent and finally

\begin{align}
&\Delta
\mathcal{T}_{TSC}^{R\pm}(p_{\parallel},q_{\parallel})=-\frac{1}{4\pi^{2}v_{F}}\ln\left(\frac{\Omega}{\omega}\right)\nonumber
\\ &\times \int_{\mathcal{D}_{2}} dk_{\parallel}\big[g_{1R}
\left(k_{\parallel},-k_{\parallel}+q_{\parallel},p_{\parallel}\right)\nonumber
\\&-g_{2R}\left(k_{\parallel},-k_{\parallel}+q_{\parallel},-p_{\parallel}+q_{\parallel}\right)\big]\mathcal{T}_{TSC}^{R\pm}(k_{\parallel},q_{\parallel}).\label{delta4}
\end{align}

\section{}

In this appendix we will present all RG equations up to two-loop for
the one-dimension case. For the quasiparticle weight we get simply

\begin{equation}
\omega\frac{d\ln
Z}{d\omega}=\frac{1}{4\pi^{2}v_{F}^{2}}\left(g_{1R}^{2}+g_{2R}^{2}-g_{1R}g_{2R}\right)\equiv\gamma
\label{z1d}
\end{equation}

\noindent In 1D the flow equations for the renormalized coupling
functions reduce simply to

\begin{equation}
\omega\frac{dg_{1R}}{d\omega}=\frac{g_{1R}^{2}}{\pi
v_{F}}+\frac{g_{1R}^{3}}{2\pi v_{F}^{2}}\label{g11d}
\end{equation}

\begin{equation}
\omega\frac{dg_{2R}}{d\omega}=\frac{g_{1R}^{2}}{2\pi
v_{F}}+\frac{g_{1R}^{3}}{4\pi v_{F}^{2}}\label{g21d}
\end{equation}

\noindent Finally, the renormalized form factors are much simpler
and they are now given by

\begin{equation}
\omega\frac{d\mathcal{T}_{CDWR}}{d\omega}=\frac{1}{2\pi
v_{F}}(2g_{1R}-g_{2R})\mathcal{T}_{CDWR}+\gamma\mathcal{T}_{CDWR}\label{cdw1d}
\end{equation}

\begin{equation}
\omega\frac{d\mathcal{T}_{SDWR}}{d\omega}=\frac{-g_{2R}}{2\pi
v_{F}}\mathcal{T}_{SDWR}+\gamma\mathcal{T}_{SDWR}\label{sdw1d}
\end{equation}

\begin{equation}
\omega\frac{d\mathcal{T}_{SSCR}}{d\omega}=\frac{1}{2\pi
v_{F}}(g_{1R}+g_{2R})\mathcal{T}_{SSCR}+\gamma\mathcal{T}_{SSCR}\label{ssc1d}
\end{equation}

\begin{equation}
\omega\frac{d\mathcal{T}_{TSCR}}{d\omega}=\frac{1}{2\pi
v_{F}}(g_{2R}-g_{1R})\mathcal{T}_{TSCR}+\gamma\mathcal{T}_{TSCR}\label{tsc1d}
\end{equation}

\noindent Consequently, the 1D resulting susceptibilities are now
determined by

\begin{equation}
\omega\frac{d\chi_{aR}}{d\omega}=\frac{1}{2\pi
v_{F}}\left(\mathcal{T}_{aR}\right)^{2}\label{qui1d}
\end{equation}

\noindent where $a=CDW,SDW,SSC,TSC$.

\end{document}